# Performance Evaluation of Flash File Systems

Pierre Olivier*, Jalil Boukhobza*, Eric Senn[+]

*Université Européenne de Bretagne
Université de Brest ;
CNRS; UMR 3192 Lab-STICC,
20 avenue Le Gorgeu
29285 Brest Cedex 3, France
`firstname.lastname@univ-brest.fr`

[+]Université Européenne de Bretagne
Université de Bretagne Sud ;
CNRS; UMR 3192 Lab-STICC,
C.R. C. Huygens
56321 Lorient, France
`eric.senn@univ-ubs.fr`

*Abstract*—Today, flash memory are strongly used in the embedded system domain. NAND flash memories are the building block of main secondary storage systems. Such memories present many benefits in terms of data density, I/O performance, shock resistance and power consumption. Nevertheless, flash does not come without constraints: the write / erase granularity asymmetry and the limited lifetime bring the need for specific management. This can be done through the operating system using dedicated Flash File Systems (FFSs).

In this document, we present general concepts about FFSs, and implementations example that are JFFS2, YAFFS2 and UBIFS, the most commonly used flash file systems. Then we give performance evaluation results for these FFSs.

*Index Terms* — Performance Evaluation, Embedded Operating System, NAND Flash Memory, File Systems.

## I. INTRODUCTION

Flash memories are divided into two main categories [1]: as NOR flash is more suitable for code execution due to low read latencies, NAND flash are commonly used for data storage. In this document we focus on NAND flash. It is today the main building block for secondary storage systems in embedded systems. This popularity is due to its many benefits: high data density, good I/O performance, shock resistance, and low power consumption. Some of these benefits are due to the fact that flash is entirely composed of electronic components, compared to hard drives containing mechanical parts.

Nevertheless, NAND flash memories (referred as "flash memory" in the rest of this document) are prone to specific constraints, due to their internal intricacies: the impossibility to perform in-place data updates, the erase / write operation asymmetry, and the limited lifetime of the memory cells. These drawbacks bring the need for specific flash management in the systems integrating such memories.

This management can be provided in a software layer by the operating system through dedicated *Flash File Systems* (FFSs).

A performance evaluation for FFS allows us to understand the behavior of such systems, and also to know how to choose the best FFS for a given hardware / software context. We can also highlight some particular points of interests in the FFS behavior, for example in order to propose optimizations.

In a first part of this document we briefly present general concepts about flash memory, and the different way it can be managed in a system. Then we focus on the use of dedicated FFSs, and we present some current implementation examples that are JFFS2, YAFFS2 and UBIFS, the most commonly used FFSs. Next we detail the main performance metrics and we give performance evaluation result comparing theses FFSs.

## II. INTRODUCTION TO FLASH MEMORIES

In NAND flash memories, data are organized in a hierarchical way: *planes* are matrices of *blocks*, which are themselves divided into *pages*. Pages contain a *user data* area, and a small metadata part called the *Out-Of-Band* (OOB) area. The blocks in recent flash memories contain between 64 and 128 pages, each page having a size varying between 2 and 8 Kilobytes (KB).

Flash memory supports 3 key operations: traditional read and write are performed at a page level. The erase operation is performed at the granularity of a block.

One of the main constraints of flash memories is the *erase-before-write* rule. Combined with the asymmetry of write and erase operations, this constraint make impossible in-place data updates. Moreover, flash memory cells can only sustain a limited number of erase operations: after a certain threshold (between $10^4$ and $10^5$) [2] they can no more retain information.

NAND flash is often shipped containing unusable bad blocks. More bad blocks appear because of wear out during the flash life cycle. Moreover, NAND flash is also unreliable, as bit errors can occur [3] during various operations.

Because of these intricacies, flash memory must be managed in a specific way when integrated in a system.

## III. FLASH MEMORY MANAGEMENT

The impossibility to perform in-place data update is bypassed by writing into another location of the flash memory, and invalidating old data location. Invalid data are recycled (erased) later by a process called the *Garbage Collector* (GC).

In order to limit the wear of the memory, and maximize its lifetime, read and erase cycles must be leveled on the whole memory surface. Flash memory management mechanisms implement *Wear Leveling* (WL) policies.

Such flash specific concepts are provided by flash management mechanisms. They can be implemented in a *hardware* way, by the use of the Flash Translation Layer (FTL) [2] in storage peripherals like USB flash drives, solid state drives or flash-based cards like SD / MMC. In embedded systems using raw flash chips, flash memory can be controlled in a software way directly by the operating system through dedicated *Flash File Systems* [1], [4], [5].

## IV. DEDICATED FLASH FILE SYSTEMS

The FFSs presented in this document are the widely used JFFS2, YAFFS2 and UBIFS. They are all integrated into the Linux kernel, officially (JFFS2, UBIFS) or through the application of a patch (YAFFS2). In the layered software of the kernel, FFSs are located on the top of the *Memory Technology Device* (MTD) layer [6]: a generic subsystem providing drivers for various memory devices. MTD allow FFSs to perform raw NAND flash access. On the top of the FFSs layer is the *Virtual File System* (VFS) Layer. VFS is a generic layer presenting directory trees from various file systems to the user with a unified way.

FFSs must assume traditional file systems functions, the storage and the indexation of a file tree. Moreover, they have to cope with flash constraints: as they perform out-of place data updates and data invalidation, they provide *garbage collector* mechanisms. They also have to implement *wear leveling* policies. The *compression* is also a feature found in many FFSs: it does not only reduce the size of stored data, but also the I/O load. *Bad block management* and *error correcting codes* are functions that must be implemented by FFSs supporting NAND flash. Bad blocks are generally identified with a marker in the OOB area, and never used. Some FFSs provide *journaling* capabilities, in order to cope with *unclean unmount* operations, for example in case of power loss.

The *Journaling Flash File System version 2* (JFFS2) [1] is today the most used FFS. It supports both NAND and NOR flash. In JFFS2, mount time and the RAM consumption are reported to scale linearly according to the managed flash size. This makes JFFS2 a poor candidate for large-sized flash memory chips. JFFS2's drawbacks lead to the development of YAFFS2 and UBIFS.

*Yet Another Flash File System* (YAFFS2) [4] is a NAND flash only FFS. It scales better than JFFS2, especially for mount time because of a technique called *checkpointing* allowing YAFFS to scan only a small part of the flash at mount time. Nevertheless, YAFFS2 still scales in a linear way.

*Unsorted Block Image File System* (UBIFS) [5] uses tree-based structures for file indexation, unlike JFFS2 and YAFFS2 which use table-based structures. Then, UBIFS scales in a logarithmic way with the size of the managed flash partition. UBIFS is then a good solution for large-sized flash partitions. UBIFS supports NAND and NOR flash.

## V. PERFORMANCE EVALUATION METHODOLOGY

Many metrics can be defined when benchmarking a FFS: traditional metrics like mount time or RAM consumption, and their evolution in relation with the partition space ; read and write I/O performance, CPU usage, tolerance to power failures, etc. In the case of FFS, it is also important to consider flash specific metrics: wear leveling, garbage collection impact on performance, and bad block management.

In this document we choose to focus on I/O and file management performance, (un)mount operations execution time, compression impact, and wear leveling management.

Regarding file manipulation, we benchmarked file tree creation and deletion performance, as well as the file search in different file trees. We generate different file trees with the following metrics: the number of files per generated directory, the number of directories per generated directory, the generated file size, and the depth of the directory tree. We wrote a tool capable of generating such directory trees, taking statistical distributions to define the first 3 metrics.

For wear leveling management, we designed a tool, Flashmon, [7] monitoring raw flash access (pages reads and write, block erasures). Performing I/O operation on various FFSs, we look at the difference between the more erased and the less erased physical block.

We also studied the impact of various compression options on the execution time of operations like file system mount, directory tree creation, file search, etc.

## VI. PERFORMANCE EVALUATION RESULTS

Results show that compression provided by UBIFS and JFFS2 considerably reduces the size of stored data, compared to YAFFS which does not provide such a feature. In terms of file manipulation, UBIFS gives the best results when creating file, as YAFFS seems to performs well when researching file metadata : YAFFS outperforms the other FFSs by a factor of two regarding *find* command execution time. UBIFS gives also good results for the mount time, as does YAFFS when using the *checkpointing* technique.